# The kinetic equation for filament density, formed during propagation of femtosecond laser radiation, in the approximation of self-consistent field

Bulygin A.D.

V.E. Zuev Institute of Atmospheric Optics, Siberian Branch of the Russian Academy of Sciences,

Academician Zuev Square, Tomsk, 634021 Russia

*Abstract: A general form of the kinetic equation (nonlinear Fokker-Planck equation) for filament number density considering the effects of their generation and decay is stated. It is consists of the phenomenological parameters, which have been determined from the direct numerical simulation of propagation of a high-power femtosecond laser pulse (HPFLP) based on the stationary nonlinear Schrödinger equation (NSE) for a series of the threshold particular cases.*

Possibility of theoretical forecast of HPFLP propagation in different regimes of multifilamentation is still unsolved problem in spite of high applied significance [1]. The cause is that the majority of the existing methods are reduced to numerical solution of *NSE* modeling HPFLP [2]. Such approach allows to study the beams with significantly less dimensions than those of the light beams applied in the atmospheric optics problem [3]. At the same time, the laws of large numbers become apparent for large multiterawatt HPFLPs [4] therefore it is natural to search the statistical regularities in multifilamentation generation [5–8]. This work is continuation of a series of the authors' papers [6, 7], which suggests a solution of a key problem: construction of the equation for the field of filament number density $n_f(r_\perp, z)$ distributed in the space of transverse coordinates $r_\perp$ and depending on the evolutionary coordinate $z$ (propagation distance of HPFLP).

### *Basic relations for Kerr medium*

Let us begin with the Schrödinger stochastic equation for the complex envelope of the light field $\tilde{E}$ [2] (random variable, generally said) in a medium of Kerr type without any sources[1]

$$\frac{\partial}{\partial z}\tilde{E}(r_\perp, z) = \frac{i}{2k_0}\frac{\partial}{\partial x^i}\frac{\partial}{\partial x_i}\tilde{E} + ik_0 n_2 \tilde{I}\tilde{E}. \qquad (1)$$

Here $k_0 = 2\pi/\lambda_0$ is the wave number ($\lambda_0 = 800$ nm); is the wave vector; $r_\perp = \{x_1, x_2\}$ and $z$ are the transverse and longitudinal coordinates, respectively; $n_2$ is the nonlinear addition to the refractive index caused by the Kerr effect ($n = n_0 + n_2 I$) [2], $\tilde{I} \equiv \frac{cn_0}{8\pi}|\tilde{E}|^2$ is light field intensity.

We introduce the Wigner function for the light field [9]:

---

[1] Further we will use the Einstein notations if it is not separately mentioned

$$J(r_\perp, s_\perp, z) \equiv \left(\frac{cn_0}{8\pi}\right)\frac{1}{2\pi}\int_{\mathbb{R}^2} \langle \tilde{E}(r_\perp + \Delta r_\perp/2, z)\tilde{E}^*(r_\perp - \Delta r_\perp/2, z)\rangle e^{-is_\perp \Delta r_\perp} d\Delta r_\perp.$$

Differentiating $z$ with respect to the Wigner function we obtain the equations of the type of the continuity equation from Eq. (1) in the approximation of its smoothness of $r_\perp$, so called the transfer equation:

$$\left(k_0 \frac{\partial}{\partial z} + s^m \frac{\partial}{\partial x^m} + 2k_0^2 n_2 \left(\frac{\partial}{\partial x^m} I\right) \frac{\partial}{\partial s_m}\right) J = 0. \tag{2}$$

Hence we derive the equations for the first two field moments $J$ по $s_\perp$ (*). Thus for zero moment corresponding to light field intensity we obtain:

$$\left(k_0 \frac{\partial}{\partial z} + \frac{\partial}{\partial x^m} \bar{s}^m\right) I = 0. \tag{3}$$

Here $\bar{s}_m \equiv S_m/I \equiv \iint_{\mathbb{R}^2} s_m J ds_1 \wedge ds_2 / \iint_{\mathbb{R}^2} J ds_1 \wedge ds_2$ is the Poynting vector normalized to intensity. In turn, we find for the Poynting vector:

$$k_0 \frac{\partial}{\partial z} S_l = -\frac{\partial}{\partial x^m} \iint_{\mathbb{R}^2} s_l s^m J ds_1 \wedge ds_2 - 2k_0^2 n_2 \iint_{\mathbb{R}^2} s_l \left(\frac{\partial}{\partial x^m} I\right) \frac{\partial}{\partial s_m} J ds_1 \wedge ds_2 = -\frac{\partial}{\partial x^m} h^m{}_l + k_0^2 n_2 \frac{\partial}{\partial x^l} I^2, \tag{4}$$

where the "velocity" correlation tensor $h_{ml} \equiv \iint_{\mathbb{R}^2} s_m s_l J ds_1 \wedge ds_2$ is introduced that corresponds to the momentum flux density tensor in hydrodynamics.

Let us consider in detail the value: $\frac{\partial}{\partial x^m} h^m{}_l$; introduce the following notation (center):

$$\theta^m{}_l \equiv h^m{}_l - \bar{s}^m \bar{s}_l I. \tag{5}$$

Further, to simplify we use reasonable physical approximations $h_{ll} \square h_{l\bar{l}}; h_{11} = h_{22}$ confirmed in a numerical simulation (see Fig. 1) for the case of multifilamentation corresponding to the situation of deterministic chaos [8]:

$$Y^{\bar{l}}{}_{[\bar{l}\underline{L}]}\Big|^{fil} \approx -\frac{\partial}{\partial x^{\underline{L}}} \theta^{\bar{l}}{}_{\bar{l}} \approx -\frac{\partial}{\partial x^l} \theta_d^{(2)} \approx -\frac{\partial}{\partial x^l} h/2. \tag{6}$$

Here $h \equiv h^l{}_l$. Then we finally obtain:

$$\left(k_0 \frac{\partial}{\partial z} + \frac{\partial}{\partial x^m} \bar{s}^m\right) \bar{s}_l = 2k_0^2 n_2 \frac{\partial}{\partial x^l} I - I^{-1}\left(\frac{\partial}{\partial x^l} \theta_d^{(2)}\right)/2. \tag{7}$$

Now pass to derivation of the equations for $\theta_d^{(2)}$ and corresponding momentum flux density tensor. Further write similarly the equation for $\theta_d^{(2)}$ from (2):

$$\left(k_0\frac{\partial}{\partial z}+\frac{\partial}{\partial x^m}\bar{s}_m^{\,\theta}\right)\theta_d^{(2)}=2k_0^{\,2}n_2\left(\bar{s}_m\frac{\partial}{\partial x_m}I^2\right), \tag{8}$$

where the vector $\bar{s}_m^{\,\theta}\equiv S^\theta_{\,m}/\theta_d^{(2)}\equiv\iint_{\square^2}(s^l s_l)s_m J ds_1\wedge ds_2 /\iint_{\square^2}(s^l s_l)J ds_1\wedge ds_2$ is introduced.

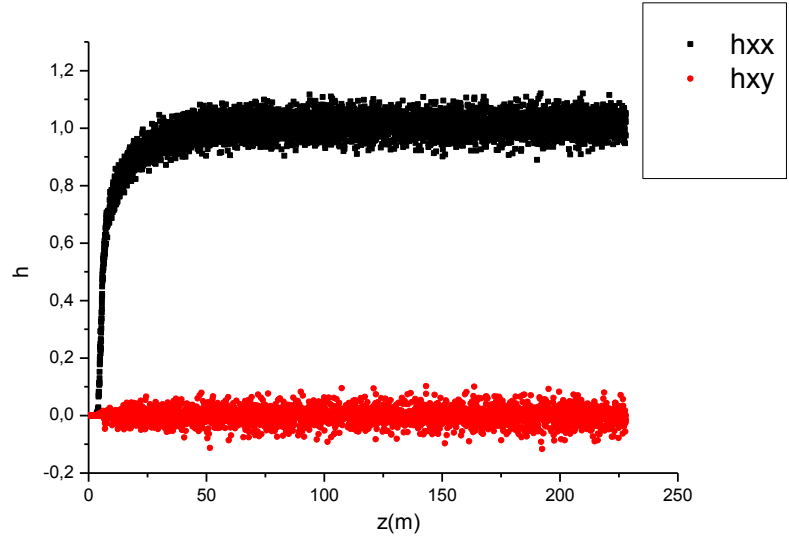

Fig. 1. "Velocity" correlation tensor $h_{xx}$ и $h_{xy}$ calculated under the model [7].

Then pass to the equation for the fluxes:

$$k_0\frac{\partial}{\partial z}S^\theta_{\,l}=-\frac{\partial}{\partial x^m}H^m_{\,l}-2k_0^{\,2}n_2\iint_{\square^2}(s^k s_k)s_l\left(\frac{\partial}{\partial x^m}I\right)\frac{\partial}{\partial s_m}J ds_1\wedge ds_2 =$$
$$-\frac{\partial}{\partial x^m}H^m_{\,l}+2k_0^{\,2}n_2\left(\frac{\partial}{\partial x^m}I\right)\iint_{\square^2}J\frac{\partial}{\partial s_m}(s_l s^k s_k)ds_1\wedge ds_2 \tag{9}$$

and consider in detail the last term:

$$\left(\frac{\partial}{\partial x^m}I\right)\iint_{\square^2}J\frac{\partial}{\partial s_m}(s_l s^k s_k)ds_1\wedge ds_2=h\frac{\partial}{\partial x^l}I+2h_l^{\,m}\frac{\partial}{\partial x^m}I \tag{10}$$

It has been determined on the ground of a direct numerical simulation that $h_{ll}=h/2$, while the terms with different indices become zero. Then

$$\left(\frac{\partial}{\partial x^m}I\right)\iint_{\square^2}J\frac{\partial}{\partial s_m}(s_l s^k s_k)ds_1\wedge ds_2=2h\frac{\partial}{\partial x^l}I \tag{11}$$

Then find from Eq. (8) neglecting the higher-order moments

$$\left(k_0\frac{\partial}{\partial z}+\frac{\partial}{\partial x_m}\bar{s}^\theta_{\,m}\right)\bar{s}^\theta_{\,l}=4k_0^{\,2}n_2\frac{\partial}{\partial x^l}I\,.$$

We finally obtain including point sources, i.e., filaments [10]:

$$\begin{cases} \left(k_0 \dfrac{\partial}{\partial z} + \dfrac{\partial}{\partial \boldsymbol{r}_\perp} \boldsymbol{s}_\perp \right) I = -\alpha_{eff} n_f \\[4pt] \left(k_0 \dfrac{\partial}{\partial z} + \boldsymbol{s}_\perp \dfrac{\partial}{\partial \boldsymbol{r}_\perp}\right) \boldsymbol{s}_\perp = 2k_0^{\,2} n_2 \dfrac{\partial}{\partial \boldsymbol{r}_\perp} I - \left(\dfrac{\partial}{\partial \boldsymbol{r}_\perp} \theta_d^{\,2}/2 - \boldsymbol{s}_\perp \alpha_{eff} n_f \right)/ I \\[4pt] \left(k_0 \dfrac{\partial}{\partial z} + \dfrac{\partial}{\partial \boldsymbol{r}_\perp} \boldsymbol{s}_\perp^{\,\theta}\right) \theta_d^{(2)} = 2k_0^{\,2} n_2 \left(\boldsymbol{s}_\perp \dfrac{\partial}{\partial \boldsymbol{r}_\perp} I^2 \right) + \gamma_{eff} n_f \\[4pt] \left(k_0 \dfrac{\partial}{\partial z} + \boldsymbol{s}_\perp^{\,\theta} \dfrac{\partial}{\partial \boldsymbol{r}_\perp}\right) \boldsymbol{s}_\perp^{\,\theta} = 4k_0^{\,2} n_2 \left(\dfrac{\partial}{\partial \boldsymbol{r}_\perp} I\right) - \boldsymbol{s}_\perp^{\,\theta}\left(2k_0^{\,2} n_2 \left(\boldsymbol{s}_\perp \dfrac{\partial}{\partial \boldsymbol{r}_\perp} I^2\right) + \gamma_{eff} n_f \right)/\theta_d^{(2)} \end{cases} \quad (12)$$

Here $\alpha_{eff}$ and $\gamma_{eff}$ are the phenomenologically introduced coefficients responsible for the dissipation effects and the effects of increment of diffraction divergence density of the light field $\bar{\theta}^2$ during filamentation, which are determined by the model of medium nonlinearity [7,10,11] (see below on page 5). Derivation of this equation for $n_f$ is based on the approximation of the self-consistent field [12,13] and also on the remark that the centers of the hot points – filaments – obey the second-order dynamic equations being in random velocity field $\tilde{\boldsymbol{s}}_\perp = \bar{\boldsymbol{s}}_\perp + \delta\tilde{\boldsymbol{s}}_\perp$. For extension of a definition of the equations, a series of the numerical simulations were required, where the empirical regularities of multifilamentation evolution in the closed system were established [7]. Write this nonlinear diffusion equation[2]:

$$\left(\dfrac{\partial}{\partial z} + \nabla_\perp \bar{\boldsymbol{s}}_\perp \right) n_f = \nabla_\perp (D \nabla_\perp n_f) - \Gamma n_f + C(\bar{I}) n_f^{3/2} - B n_f^{\,2} + F_{out}(\boldsymbol{r}_\perp, z) . \qquad (13)$$

Here the diffusion coefficient $D$ has simple relationship with diffraction divergence density: $D \approx L_m \theta^2(z, \boldsymbol{R}_\perp)/k_0^{\,2}$, it is consequence of standard definition of the diffusion coefficient $D(\boldsymbol{r}_\perp, z)\delta_{L_m}(z - z') = \bar{\theta}_d^{\,2}(\boldsymbol{r}_\perp, z)/k_0^{\,2}$ and is caused first of all by random component of "velocity" field [3] $\bar{\theta}_d^{\,2}(\boldsymbol{r}_\perp, z) = \langle \delta\tilde{s}(z) \delta\tilde{s}(z') \rangle$, where $L_m$ is memory length (length of longitudinal correlation of amplitude-phase perturbations (rings) induced by filaments), which has been found on the basis of the numerical simulation under the stationary model [7] and is on the order of ten centimeters; $\delta_{L_m}$ is regularized delta function with regularization parameter $L_m$. Other quantities including in Eq. (13) have the following values: $\Gamma$ is filaments decay "velocity" [12], i.e., the value inverse to filaments length being for air about 4–6 m; $C(\bar{I}) = C_0 + C_1 \bar{I}$ is the coefficient responsible for multiplication of the filaments [12] owing to interference of the rings of the filaments including interference with the background field [11] (coefficient $C_1 \bar{I}$); and the coefficient $B$ providing existence of the finite solutions of Eq. (13) for any $\bar{I}$ and corresponding to the saturation effects [11]; the inhomogeneous term $F_{out}(\mathbf{r}_\perp, z)$ is responsible for external mechanisms of system excitation, such as initial conditions and medium turbulence.

---

[2] It should be noted that this equation can be rewritten in the form of the Fokker-Planck equation adding $\nabla_\perp (n_f \nabla_\perp D)$ to the right and left parts and introducing $\boldsymbol{s}_\perp^D \equiv \boldsymbol{s}_\perp + \nabla_\perp D$.

[3] This circumstance is caused by the fact that conversions of amplitude perturbations into phase ones are implemented rapidly, i.e., at the distances considerably less not only than scales of variation of macroscopic characteristics of the system, but also the scales of filament length [14].

Let us demonstrate briefly our reasoning, on the basis of which the right part of the equation for $n_f$ has been written. It is found from the numerical simulations for the case $\nabla_\perp n_f = 0$ under the conditions of establishment of equilibrium for the closed system that the amplitude of filament number density $A_n^{eq} \equiv \sqrt{n_f^{eq}}$ increases according to linear law (Fig. 2) from the mean intensity $\bar{I}$ after achievement threshold level of the mean intensity $\bar{I}_{cr} \approx (2 \div 2.5) \cdot 10^{13}$ W/cm$^2$ [7]:

$$A_n^{eq}(\bar{I}) \approx \alpha_0 + \alpha_1 \bar{I}. \tag{14}$$

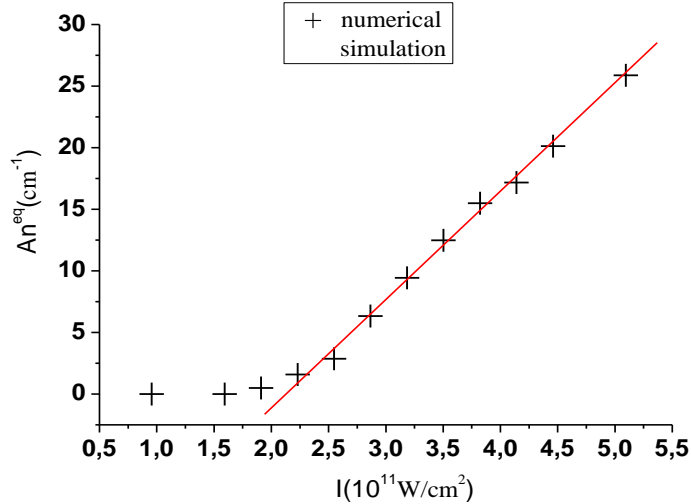

Fig. 2. Dependence of filament number density $A_n^{eq} \equiv \sqrt{n_f^{eq}}$ in equilibrium state on the mean intensity of the light field $\bar{I}$ found in the numerical simulation (crosses) and extrapolation of this dependence after critical area $\bar{I} > \bar{I}_{cr}$ (red line).

This result was obtained on the basis of the nonlinear Schrödinger equation (NSE) in the conservative form ($\alpha_{eff} = 0$):

$$\frac{\partial}{\partial z} U(\mathbf{r}_\perp, z) = \left( \frac{i}{2n_0 k_0} \nabla_\perp^2 + i k_0 \Delta n_{eff} I \right) U(\mathbf{r}_\perp, z) \tag{15}$$

In our current modeling scheme, the negative contribution of the plasma involved in a full three-dimensional model is replaced by the effective negative refractive index of the following form:

$$\Delta n_{eff}(I/I_f, \chi) = -n_2 \left( (1 + I/I_f) e^{-\chi I/I_f} - I/I_f \right)$$

Here $\chi$, $I_f$ are the empirical parameters determined from our previous work [7], which gives rise to a clamped intensity of $5\cdot10^{13}$ W/cm$^2$ and the value $\gamma_{eff}$ corresponded to a similar size in the complete model. The numerical simulation of solution of Eq. (15) was performed using the TSU SKIF Cyberia cluster. It should be noted that the approach of choice of the effective stationary model of nonlinearity is similar to those suggested in [11,15], but comparing with them, as our numerical experiments have shown, it allows to adjust more flexibly formation of angular divergence of the light beam to the complete model.

Further it should be noted that the abovementioned characteristic of appearance of the critical parameter is typical for the nonlinear systems [16], phase portrait for which has the following form (Fig. 3).

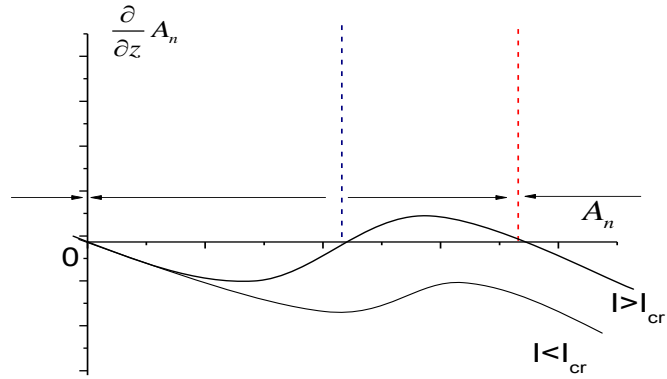

Fig. 3. Schematic phase portrait of the equation for the filament number density amplitude. Two phase curves are related to two values of the mean intensity parameter. Two additional equilibrium points (except trivial point at zero value of $A_n$) are formed, when mean intensity increases critical value: stable (red dotted line) and unstable (blue dotted line). Horizontal arrows show the directions of phase movement of the system.

The following dynamic equation for $A_n$ corresponds to this phase portrait:

$$2\frac{d}{\Gamma dz}A_n = -A_n + C_\Gamma(\bar{I})A_n^2 - B_\Gamma A_n^3. \qquad (16)$$

Here $C_\Gamma \equiv C/\Gamma$; $B_\Gamma = B/\Gamma$. Physical meaning of the coefficients of $A_n$ has been explained above, however the values of $C_\Gamma(\bar{I})$ and $B_\Gamma$ are not found yet. For extension of their definition we choose maximally simple type of dependence of these coefficients on mean intensity conforming this to the known fact that filament generation is connected with interference with the background field: $C_\Gamma(\bar{I}) = C_\Gamma^{(0)} + C_\Gamma^{(1)}\bar{I}$ [11]. Moreover, we note that the right part of Eq. (16) has two nontrivial roots besides trivial one; the largest from them is related to the stable equilibrium point and this is nothing else but $A_n^{eq}(\bar{I})$. Now we require that dependence of this maximal root of Eq. (16) on $\bar{I}$ satisfies empirically found dependence, then we obtain: $B_\Gamma = 1/4\alpha_0^2$; $C_\Gamma^{(0)} = 1/4\alpha_0$; $C_\Gamma^{(1)} = B_\Gamma \alpha_1$. Thus we have found a type of the self-action term responsible for generation and decay of filaments of Eq. (13) and suppose the procedure of their specification from the numerical simulation.

In conclusion we make a few remarks about the limits of applicability of the system of Eqs. (12)–(13) and form of function of external source $F_{out}(\boldsymbol{r}_\perp, z)$.

The equations for filament number density have been obtained in the approximation of energy constancy in an area element, therefore they can be correct only under the conditions of relative smoothness of mean intensity, or at fulfillment of the following condition: $\nabla_\perp (\bar{s}_I \bar{I}) \ll \Gamma$. With respect to the source function $F_{out}(\boldsymbol{r}_\perp, z)$ in the presence of constant external disturbances, such as medium turbulence, the independent investigation should be done, which we are going to perform in following works. In the case of absence of turbulence, the analytical construction of the function $F_{out}(\boldsymbol{r}_\perp, z)$ is possible only for the simplest situation, when each disturbance on the intensity profile evolves rather independently; and the beginning of its filamentation can be found by the Marburger equation [2]. In general case, obtaining of this function is possible only by direct numerical modeling of NSE at the initial stage of propagation of HPFLP with subsequent joining with the equation system (12)–(13).